\documentclass[10pt]{article}
\usepackage{epsfig}
\usepackage{natbib}
\usepackage[dvips]{color}

\newcommand{\bq}{\begin{equation}}
\newcommand{\eq}{\end{equation}}
\newcommand{\dg}{$^\circ$}

\newcommand{\as}{$''$}

\newcommand{\seven}{$J=7\rightarrow6$}
\newcommand{\eight}{$J=8\rightarrow7$}
\newcommand{\nine}{$J=9\rightarrow8$}
\newcommand{\eleven}{$J=11\rightarrow10$}

\newcommand{\Tas}{$T^*_A$}
\newcommand{\Tr}{$T_{rx}$}
\newcommand{\Ta}{$T_A$}
\newcommand{\Tm}{$T_{atm}$}

\newcommand{\coTW}{$^{12}$CO}
\newcommand{\coTH}{$^{13}$CO}

\title{\bf Characterization and Status of a Terahertz Telescope}
\author{\small Daniel P. Marrone, Raymond Blundell, Hugh
Gibson\footnote{Present address: RPG Radiometer Physics GmBH}, Scott
Paine, D. Cosmo Papa, C.-Y. Edward Tong\\
\small\it Harvard-Smithsonian Center for Astrophysics\\
\small\it 60 Garden Street, Cambridge, MA 02138 USA\\
\small\it dmarrone@cfa.harvard.edu}
\date{}

\begin{document}
\maketitle
\thispagestyle{empty}

\begin{abstract}
\normalsize{The Receiver Lab Telescope (RLT) is a ground-based
terahertz observatory, located at an altitude of 5525 m on Cerro
Sairecabur, Chile. The RLT has been in operation since late 2002,
producing the first well-calibrated astronomical data from the ground
at frequencies above 1~THz. We discuss the status of this telescope
after 18 months of operation and plans for the upcoming observing
season.

There are many practical challenges to operating a telescope at these
frequencies, including difficulties in determining the pointing,
measuring the telescope beam and efficiency, and calibrating data,
resulting from high receiver noise, receiver gain instabilities, and
low atmospheric transmission. We present some of the techniques we
have employed for the RLT, including the use of atmospheric absorption
lines in the place of continuum measurements for efficiency and beam
measurements, and the utility of a Fourier-transform spectrometer for
producing reliable data calibration.}

\end{abstract}

\section*{\normalsize 1 Introduction}
\begin{figure}[t]
\begin{center}
\epsfig{file=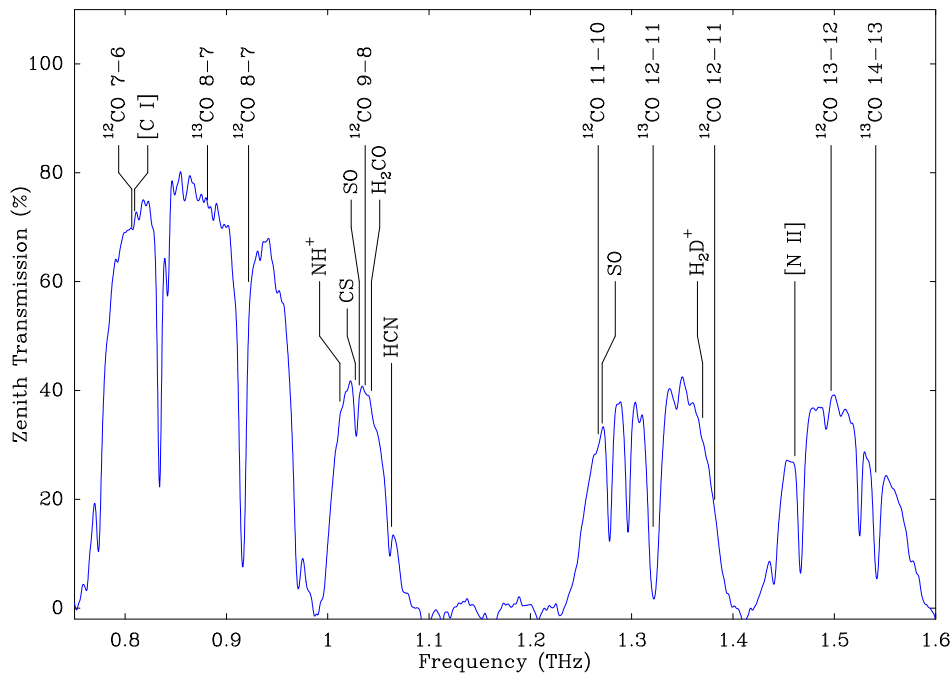,angle=270,width=125mm}
\end{center}
\vspace{-5mm}
\caption{An example of the terahertz windows available from Cerro
Sairecabur. This atmospheric transmission spectrum is actual data from
the Receiver Lab Fourier-transform spectrometer (discussed briefly in
\S3.1, see also Paine et al. 2000), obtained on a very good day at the
RLT site. The frequencies of several astronomically interesting lines
are also plotted.}
\label{f-tx}
\end{figure}

Astronomy at frequencies above 1~THz has long been considered
impossible from ground-based observatories because of strong
atmospheric absorption. As a result, there is a great deal of effort
currently directed toward developing airborne and space observatories
for this portion of the electromagnetic spectrum. Development of the
HIFI instrument of the Herschel satellite is discussed extensively in
these proceedings. Recently it has been shown that terahertz astronomy
is possible from extremely dry ground sites, such as the South Pole
and high in the Atacama Desert of Chile, where transmission is
routinely observed in several windows between 1 and 3~THz (Paine et
al. 2000, Matsushita et al. 2000). Many groups are now attempting to
use the windows offering the highest transmission, centered at 1.03,
1.35, and 1.5~THz (see Figure~\ref{f-tx}), for astronomical
observations. The Antarctic Submillimeter Telescope and Remote
Observatory (AST/RO) (Stark et al. 2001) now has two instruments in
place for such observations: SPIFFI (Nikola, et al. 2004), and TREND
(Gerecht et al. 2003). The 12-meter APEX telescope, located at the
ALMA site, will also have receivers for these terahertz windows in the
next year.

The first (and currently only) telescope to make routine ground-based
observations in these terahertz windows is the Receiver Lab Telescope,
built by the Smithsonian Astrophysical Observatory (SAO) in
collaboration with the Universidad de Chile. The development of this
telescope has been reported on in previous editions of these
proceedings, beginning with its introduction by Blundell et
al. (2002), and followed by the presentation of early data (Radford et
al. 2003). In this paper we provide a summary of the status of the RLT
and planned upgrades (\S2). The unique challenges of ground-based
terahertz observations are discussed in \S3. In order to properly
calibrate our data we have explored new techniques (\S3.1, 3.2) which
may be of interest at other observatories facing similar conditions.

\section*{\normalsize 2 The Receiver Lab Telescope}
The RLT is located on Cerro Sairecabur in northern Chile, 40~km north
of the future ALMA site. This site, at a latitude of 22.5{\dg}~S, is
well suited for studies of the inner galaxy and nearby spiral arms as
most objects in these regions pass almost directly overhead. The
telescope sits at an elevation of 5525~meters, 500~meters higher than
ALMA, making it the highest telescope in the world. During the SAO
site testing program that preceded the construction of the RLT, the
Receiver Lab Fourier-transform spectrometer (FTS) was used at the ALMA
site (Paine et al. 2000) and on Sairecabur. It was found that the
Sairecabur site shows better peak transmission and a higher fraction
of time available for terahertz observations because it is more often
above the atmospheric inversion layer responsible for trapping
moisture. An example of the transmission at Sairecabur is shown in
Figure~\ref{f-tx}.

\begin{figure}[t]
\begin{center}
\epsfig{file=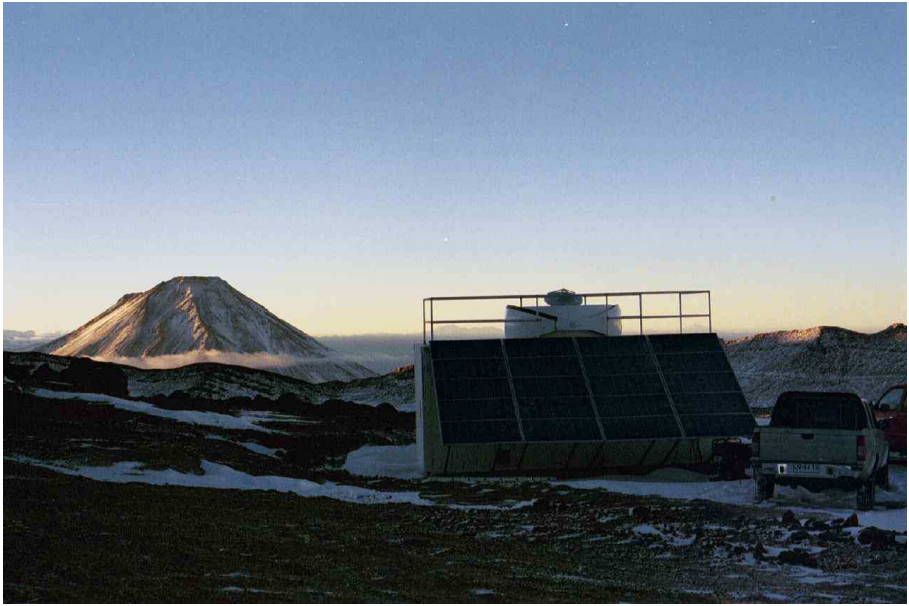,width=120mm}
\end{center}
\vspace{-5mm}
\caption{The Receiver Lab Telescope, on Cerro Sairecabur, Chile.}
\label{f-rlt}
\end{figure}

The RLT (Figure~\ref{f-rlt}) is an 80~cm alt-az telescope sitting atop
a 20-foot shipping container. The container houses the telescope
tower, receivers, correlator, control computer, and equipment
associated with the solar power system. The telescope is outfitted
with two receivers for frequencies between 800~GHz and
1.6~THz. Frequency coverage is currently limited by the availability
of solid state local oscillators, so the 1.5~THz window is not
accessible. The mixers are waveguide-mounted phonon-cooled HEBs. All
of those used on the telescope have been NbN devices, with and without
MgO buffer layers. Typical noise temperatures are 950-2000~K in our
frequency range. The 1~GHz wide IF is centered on 3~GHz. Spectra are
obtained from a 330 channel digital autocorrelation spectrometer with
3~MHz resolution (1~km/s at 1~THz).

The RLT routinely observes spectral lines that are rarely, if ever,
observed anywhere else. The most commonly used lines are {\coTW}
{\seven} (806.65~GHz), {\coTH} {\eight} (881.27~GHz), and {\coTW}
{\nine} (1.03691~THz). With an 85{\as} beam at 1~THz, the RLT is well
suited to mapping of extended emission, such as that arising in large
H II regions. The first RLT terahertz spectrum was obtained on
November 11, 2002. First scientific results can be seen in Marrone et
al. (2004).

In the near future there are many upgrades planned for the
telescope. In May we will begin observations of the CO {\eleven} line
at 1.2670~THz, the first ground-based observations of this
line. Expansion to the 1.5~THz window awaits a local oscillator, which
should arrive this year. The correlator is due to be replaced with a
new 512-channel correlator with 1 and 2~MHz resolution, providing
greater detail in the narrow galactic emission lines we normally
observe.

\section*{\normalsize 3 Challenges of Ground-based Terahertz Astronomy}
Ground-based terahertz telescopes have some advantages over airborne
or spaceborne instruments. Ground telescopes can be made larger than is
feasible in the air or in space, are considerably cheaper, are easier
to service than satellites, and may provide more observing time than
can be obtained from an aircraft. Of course, observations from the
ground are strongly limited by the properties of the atmosphere. Most
importantly, observations are only possible in a few windows, but even
within these windows the high atmospheric opacity, lack of strong
calibrators, and the instabilities of terahertz receivers conspire to
make common pointing, calibration, and telescope characterization
tasks difficult. In the following sections we discuss solutions to
these problems derived from 18 months of RLT operations.

\subsection*{\normalsize 3.1 Atmospheric Calibration}
In order for spectra obtained from a terahertz telescope to be
scientifically useful they must be properly calibrated. A common
convention is to correct measured spectra for receiver gain,
inefficiencies in the telescope, and atmospheric absorption, placing
the observed line on a temperature scale that would be observed by a
perfect telescope above the atmosphere (this is the {\Tas} scale of
Ulich \& Haas 1976). The atmospheric correction, and typically the
efficiency correction, requires a reliable method of determining the
opacity.

At the best well-characterized observing sites the atmospheric
transmission in the windows above 1~THz is no better than $\sim40$\%,
and is generally much lower. In addition, our experience at Cerro
Sairecabur shows that large variations in transmission are common: in
a single night the zenith transmission can vary by as much as 50\% of
its peak value. Changes in transmission between and within observing
nights, whether due to changing source elevation or atmospheric
fluctuations, can have a significant effect on the apparent strength
of a spectral line. Without reproducible atmospheric correction, data
taken on different nights, or even at different times in the same
night, cannot be usefully compared.

There are several techniques for properly accounting for the
atmospheric attenuation. Most radio astronomy calibration techniques
have been developed at lower frequencies where the atmospheric effects
are smaller and receivers are more sensitive and stable, making it
difficult to apply them directly to observations with the RLT. One
such example is the skydip, in which the sky temperature is measured
as a function of elevation and the resulting curve is used to
determine the mean atmospheric temperature and zenith
opacity. Although the variation in sky temperature with elevation is
small when the atmospheric transmission is low, it is still easily
measurable with RLT receivers. The most significant problem we have
encountered with this technique is that instabilities in the power
output of the receiver are greatly magnified in skydips when
atmospheric transmission is low.

The magnification of power instabilities by skydips can be understood
from the equation for the antenna temperature (\Ta) on blank sky. At
elevation $\epsilon$, with receiver noise {\Tr}, mean atmospheric
temperature {\Tm}, and zenith transmission $t$, the measured {\Ta}
is:
\begin{eqnarray}
T_A = T_{rx} + T_{sky} + T_{cmb} \hspace{4.3pt} \nonumber \\
 \simeq T_{rx} + T_{atm}\left(1-t^x\right)
\label{e-ta-sky}
\end{eqnarray}

\noindent where $x = csc(\epsilon)$. The cosmic background term
($T_{cmb}$) has been dropped in the second line because the
Rayleigh-Jeans brightness temperature of the CMB is vanishingly small
at 1~THz. The transmission inferred from {\Ta} measured at one
elevation (as a function of the usually unknown \Tm) can be found by
rearranging equation~\ref{e-ta-sky}:
\bq
t = \left(1 - \frac{T_A -
T_{rx}}{T_{atm}}\right)^{1/x}
\label{e-tx}
\eq

\noindent Instabilities in the IF output power ($P_A$), whether they
originate in the receiver or the atmosphere, correspond to changes in
the measured {\Ta} because \Ta $=K P_A$, where $K$ is a conversion
factor between power and antenna temperature determined using hot and
cold loads. Differentiating equation~\ref{e-tx} with respect to $P_A$
and substituting for $T_{sky}$ (as in equation~\ref{e-ta-sky}), we
obtain:
\bq
\frac{\partial t}{\partial P_A} = \frac{-t}{x P_A}\left[
\left(1+\frac{T_{rx}}{T_{atm}}\right)t^{-x} - 1\right]
\eq

\noindent Ignoring all other error contributions, we can translate
this equation into a relationship between the rms power fluctuations
and the rms error on $t$. By noting that {\Tas} is proportional to
$t^{-x}$, we can go directly to the fractional error
($\sigma_{T^*_A}/T^*_A$) in spectra calibrated using
a single skydip transmission measurement for power fluctuations
$\sigma_{P_A}/P_A$:
\bq
\frac{\sigma_{T^*_A}}{T^*_A} = \left[
\left(1+\frac{T_{rx}}{T_{atm}}\right)t^{-x} - 1
\right] \frac{\sigma_{P_A}}{P_A}
\label{e-skydip}
\eq

\noindent In reality, one does not use a single measurement to
determine both $t$ and {\Tm}. For a more typical 10 point skydip,
using RLT values for {\Tr} (1000~K), {\Tm} (250~K), and $t$ for
frequencies above 1~THz (20\%), the random calibration error
contributed by transmission measurement error is of the order of 15-20
times the power instabilities. A typical rms noise on the continuum
power output of an RLT receiver is 0.5\%, making this a significant
cause of error in calibration.

\begin{figure}[t]
\begin{center}
\epsfig{file=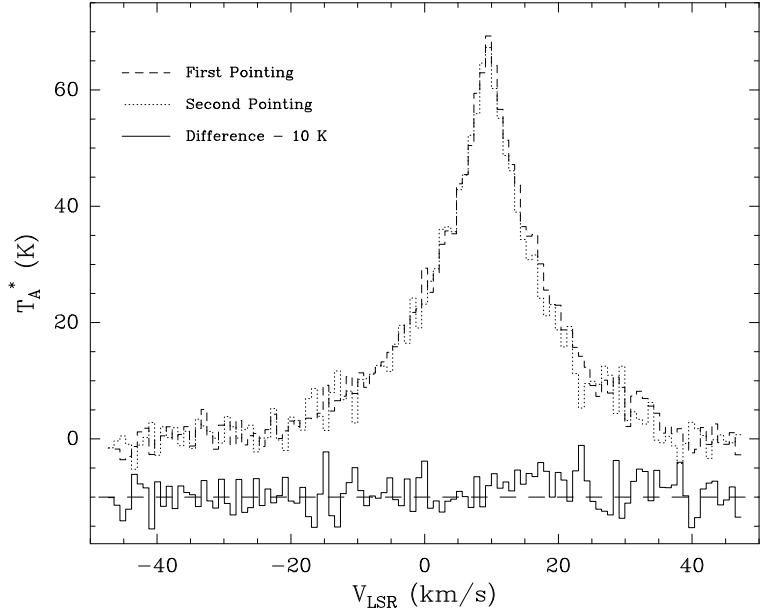,angle=270,width=100mm}
\end{center}
\vspace{-5mm}
\caption{An example of the results of our FTS calibration
technique. Shown are two spectra of Orion KL in CO {\nine}
(1.037~THz), each required 1 minute of on-source integration time and
the two observations were made one hour apart. The system temperature
increased by 56\% between the observations, but the difference between
the two calibrated spectra is almost consistent with the rms noise
seen in the baseline channels.}
\label{f-calibration}
\end{figure}

Rather than the skydip method described here, we directly measure the
zenith transmission using our FTS during all RLT observations. This
instrument generates an atmospheric transmission spectrum from 300 to
3500~GHz at 3~GHz resolution every 10 minutes, and does not require
any of the telescope observing time (as skydips would). An example of
the reproducibility of the FTS calibration is shown in
Figure~\ref{f-calibration}. Despite an interval of an hour between two
observations of the same object, during which time the transmission
was observed to decrease from 22.5\% to 19\% (or 19\% to 12.5\% at
the source elevation), the two calibrated spectra show very similar
amplitude. More details about the use of an FTS in calibrating
astronomical data, including refinements that can be made using
atmospheric models, can be found in Paine \& Blundell 2004.

\subsection*{\normalsize 3.2 Telescope Characterization}
The RLT has also faced difficulties in determining the telescope
efficiency, beamshape, and radio pointing. Typically, these properties
would all be measured by mapping the continuum emission of a planet,
but we have found that drifts in the power output of our HEB receivers
make continuum maps unreliable. 

An alternative approach is to use a more narrowband signal as a
calibration source, since other noise sources should be less important
relative to radiometric noise in a smaller bandwidth (e.g. Schieder \&
Kramer 2001). Narrow, unsaturated atmospheric lines, such as O$_3$
lines, provide exactly such a signal when observed against the disk of
a planet. Moreover, by using planets as a backlight we retain the
source geometry that is normally used for these purposes.

Our procedure for obtaining a beam map, pointing offsets, and a beam
efficiency is as follows. We make a map of a planet using an ozone
line near to the observing frequency of interest. Once calibrated, we
fit each spectrum in the map with a model ozone line profile to
determine a best-fit line depth. The line depths form the basis for a
beam map because the variations in line depth reflect changes in the
coupling of the telescope beam to the planet (rather than an
atmospheric effect), just as continuum measurements would. By fitting
the derived map we obtain a beamshape and an offset from the nominal
source position. The fit also yields the line depth that would be
observed if there were no pointing error, which, after accounting for
the coupling of the beam to the planet, can be used to determine the
efficiency.

Our method and the standard continuum method are similar in many
respects. We require good calibration to ensure that the individual
map spectra can be reliably compared to each other, but this is no
different than what is required for a continuum map taken over a long
enough interval that the atmospheric transmission and source elevation
change. Once the calibrated line depths are obtained, the procedures
for determining the pointing and beam shape are the same as one would
use with a continuum map. The only additional requirement of this
portion of our method is a model ozone lineshape, which we obtain from
the {\it am} atmospheric model\footnote{{\it am} is available at:
http://cfarx6.harvard.edu/am} (Paine 2004) and a vertical ozone
profile from ozonesonde and lidar observations made from Hawaii
(Oltmans et al. 1996, Leblanc \& McDermid 2000). The line depths we
determine are only weakly dependent on the assumed vertical ozone
profile for reasonable profiles. For the efficiency determination we
need a good estimate of the true strength of the ozone absorption. The
absorption represents the known signal to which we compare our
observed signal and obtain an efficiency, analogous to comparing a
measured continuum brightness temperature to an expected planetary
brightness temperature. Fortunately, the total O$_3$ column is
measured daily at many sites worldwide, including the nearby
Marcapomacocha, Peru, and is made available by the NOAA Climate
Monitoring and Diagnostics Laboratory\footnote{Ozone column
measurements are available on-line at:
http://www.cmdl.noaa.gov/ozwv/dobson}.

\begin{figure}[t]
\begin{center}
\epsfig{file=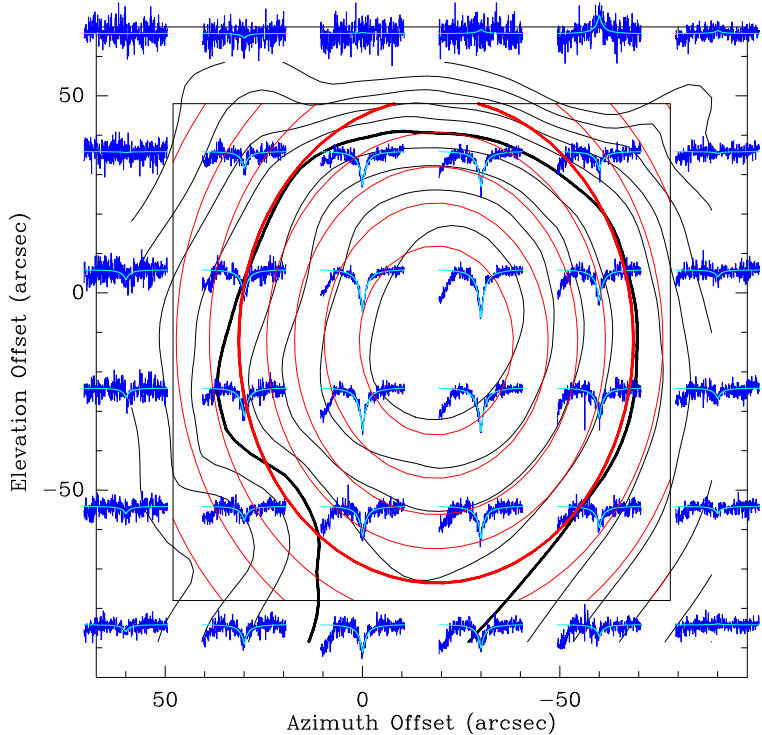,angle=270,width=100mm}
\end{center}
\vspace{-5mm}
\caption{An O$_3$ beam map made with the RLT against Jupiter. Spectra
were taken on a 30{\as} grid in an irregular order to reduce the
biasing effects of calibration errors. The individual spectra
({\color{blue} \bf -----}) and best fit line profiles ({\color{cyan}
\bf -----}) are shown at each map point. The absorption at the left edge
of each spectrum is a correlator artifact. The beam is fit to the
central 16 spectra (enclosed in a box). The measured ({\bf -----}) and
best fit ({\color{red} \bf -----}) beams are contoured in 10\% steps
with the 50\% contour reinforced. The two contour sets are seen to
agree well over most of the map, except toward the north where noisy
spectra that have been excluded from the fit cause the measured
contours to deviate from the best fit contours.}
\label{f-beammap}
\end{figure}

Figure~\ref{f-beammap} is an example of a beam map made with the
technique just described. The map was made using the 883.057~GHz line
of ozone, with Jupiter (41.7{\as} diameter) as the continuum
source. From this map we find pointing offsets of -19.0$\pm$1.6{\as}
and -12.4$\pm$2.1{\as} and FWHM beamsizes of 96$\pm$5{\as} and
117$\pm$9{\as} in azimuth and elevation, respectively, along with a
line depth of 5.3$\pm$0.3~K\footnote{These errors are obtained from
simulations of the effects of reasonable random errors in the zenith
transmission and receiver noise, and will be larger when systematic
effects are included.}, which corresponds to an efficiency of
44\%. The measured pointing offsets are small compared to the beamsize
and are comparable to those suggested by previous terahertz
observations. The beamsize is similar to what is expected from the
optics design, while the efficiency suggests that at the time that
this map was made the receiver was shifted slightly from the designed
position so that its beam was clipped before reaching the telescope.

We believe that the use of ozone lines as a calibration tool can also
be extended to other telescopes. To determine whether this is a useful
technique at a given telescope one must compare the fluctuations in
the receiver output power, scaled by the system temperature, to the
signal strength from available planets. For the RLT and Jupiter, the
expected signal is around 10~K at 1~THz, but the receiver
instabilities (of the order of 0.5\% rms in 1 second integrations on
the RLT) applied to a $T_{sys}$ of 20000~K correspond to 100~K
fluctuations. Larger telescopes see a larger planetary signal, so they
may be able to tolerate small instabilities. On the other hand, if
beam maps are made in poorer transmission or at lower source elevation
$T_{sys}$ can be significantly larger than the 20000~K used here. The
use of ozone as a calibrator is not something that restricts this
technique to Chile; daily ozone measurements are also made from Mauna
Loa, Hawaii, and the South Pole, so this information is available for
all major submillimeter sites. This technique can be used near to any
frequency of interest because suitable ozone lines are available
throughout all of the terahertz windows. As a final consideration, the
measurement technique employed in making the beam map may affect the
expected depth of the ozone line. For position- or beam-switched
observations, where the map spectra are obtained by combining on- and
off-source spectra as (ON-OFF)/OFF, the expected line depth is not
simply proportional to the planet signal times the fractional
absorption due to ozone. As {\Tr} decreases, a term in the expansion
of the line depth that is insignificant for the RLT ($<$5\% for
$T_{rx}+T_{sky}=1200$~K) grows like $1/\left(T_{rx}+T_{sky}\right)$
and could be relevant for lower noise terahertz receivers.

\section*{\normalsize 4 Conclusions}
We have discussed the status of the SAO Receiver Lab Telescope,
presently the only telescope making ground-based astronomical
observations at frequencies above 1~THz. We are currently beginning
observations in the 1.3~THz atmospheric window, while observations in
the 1.5~THz window await a local oscillator. Neither of these windows
has previously been used for astronomical observations.

We have also presented techniques for making ground-based terahertz
observations. In particular, we have discussed the difficulty of using
skydips for data calibration and the power of a Fourier-transform
spectrometer as a calibration tool. We have also presented a new
technique for characterizing a telescope in the presence of high
system temperatures and receiver instabilities. 

\section*{\normalsize References}

\noindent Blundell, R. et al. 2002, in Proc. Thirteenth International
Symposium on Space Terahertz Technology\\

\noindent Gerecht, E. et al. 2003, in Proc. Fourteenth International
Symposium on Space Terahertz Technology\\

\noindent Leblanc, T. \& McDermid, I. S. 2000, Journal of Geophysical
Research, 105, 14613 \\

\noindent Marrone, D. P., et al. 2004, Astrophysical Journal in press
(astro-ph/0405530) \\

\noindent Matsushita, S., et al. 2000, in Proc. SPIE Vol. 4015, Radio
Telescopes, ed. H. R. Butcher, 378-389\\

\noindent Nikola, T, et al. 2004, these proceedings\\

\noindent Oltmans, S. J. et al. 1996, Journal of Geophysical Research,
101, 14569 \\

\noindent Paine, S. et al. 2000, Publications of the Astronomical
Society of the Pacific, 112, 108\\

\noindent Paine, S.  2004, ``The am Atmospheric Model'' SMA Technical
Memo \#152, available at \\
\indent http://sma-www.cfa.harvard.edu/private/memos\\

\noindent Paine, S. \& Blundell, R. 2004, these proceedings\\

\noindent Radford, S. J. E., et al. 2003, in Proc. Fourteenth
International Symposium on Space Terahertz Technology\\

\noindent Schieder, R. \& Kramer, C. 2001, Astronomy \& Astrophysics,
373, 746 \\

\noindent Stark, A. A. et al. 2001, Publications of the Astronomical
Society of the Pacific, 113, 567\\

\noindent Ulich, B. L. \& Haas, R. W. 1976, Astrophysical Journal
Supplement Series, 30, 247 \\

\end{document}